\documentclass[manuscript,screen]{acmart}
\AtBeginDocument{%
  }

\usepackage{algorithmic}
\usepackage{algorithm}
\usepackage{tabularx}

\usepackage{amssymb}
\usepackage{pifont}

\newcommand{\xmark}{\ding{55}}

\usepackage{wasysym}
\setcopyright{acmlicensed}
\copyrightyear{2026}
\acmYear{2026}
\acmDOI{XXXXXXX.XXXXXXX}
\acmConference[Conference acronym 'XX]{Make sure to enter the correct
  conference title from your rights confirmation email}{June 03--05,
  2018}{Woodstock, NY}
\acmISBN{978-1-4503-XXXX-X/2018/06}

\begin{document}

\title[Agentic Autoscaling for LLM-driven Text Classification]{Agentic Autoscaling through Worker-Pool Orchestration for LLM-driven Text Classification in Cloud Computing Environments}


\author{Bablu Kumar}
\email{bablu.kumar.1@student.unimelb.edu.au}
\author{Anshul Verma}
\email{anshul.verma@unimelb.edu.au; anshul.verma@bhu.ac.in}
\affiliation{%
  \institution{the Quantum Cloud Computing and Distributed Systems (qCLOUDS) Laboratory, School of Computing and Information Systems, The University of Melbourne, Australia and the Department of Computer Science, Banaras Hindu University}
  \city{Varanasi}
  \country{India}}
  \correspondingauthor
  \authornotemark[1]

\author{Rajkumar Buyya}
\email{rbuyya@unimelb.edu.au}
\affiliation{%
  \institution{the Quantum Cloud Computing and Distributed Systems (qCLOUDS) Laboratory, School of Computing and Information Systems, The University of Melbourne}
 \city{Melbourne}
  \country{Australia}}

\renewcommand{\shortauthors}{B. Kumar, A. Verma, R. Buyya}

\begin{abstract}
  The growing adoption of large language model (LLM)-based systems for large-scale text processing has created a critical need for dynamic autoscaling to manage high-latency, bursty, and computationally intensive workloads. This paper proposes an agentic autoscaling framework through worker-pool orchestration for LLM-driven text classification. The framework integrates a priority task queue, a dynamic pool of agent workers, a real-time metrics collector, and an application-layer autoscaler. Its classifier-agnostic design supports both zero-shot and fine-tuned language models without modifying the autoscaling logic. The framework is evaluated using Autoscaling+BART and Autoscaling+DeBERTa against static allocation and standalone RoBERTa and DistilBERT baselines. On the AG News dataset, Autoscaling+BART achieves 84.5\% accuracy, while Autoscaling+DeBERTa improves it to 90.5\%. On the SMS Spam Collection dataset, Autoscaling+DeBERTa achieves 99.5\% accuracy, whereas Autoscaling+BART attains 84.5\% accuracy with lower execution time. Overall, the proposed framework consistently outperforms the baseline approaches in resource efficiency while maintaining high classification performance, demonstrating that elastic worker-pool orchestration provides an effective and cost-efficient solution for scalable LLM-driven text classification in cloud environments.
\end{abstract}



\maketitle

\section{Introduction}
The proliferation of large language models (LLMs) as autonomous processing units, commonly referred to as agents, has enabled the automation of data-intensive natural language processing tasks such as text classification, summarization, and information extraction~\cite{chang2024survey,yang2024give,porter2025insyte}. An agent autonomously receives a task, invokes a language model, and returns a result without human intervention~\cite{chowa2026language,porter2025insyte}. As organizations increasingly adopt agentic pipelines for applications including customer support, news analysis, spam filtering, and content moderation, efficient resource provisioning has become a critical challenge~\cite{wang2024survey,7937885,yuan2026adaptive}. Unlike conventional web services, LLM-driven agentic workloads exhibit high inference latency, bursty task arrivals, and significant computational cost, particularly when using GPU-backed infrastructure or commercial inference APIs~\cite{cai2024llmaas,7937885}. Consequently, traditional web-service autoscalers designed around infrastructure-level metrics are often inadequate for these workloads, motivating the need for autoscaling mechanisms specifically designed for agentic systems.

A straightforward solution to this provisioning problem is static resource allocation, where a fixed number of worker processes is provisioned according to either average or peak workload. Provisioning for average demand leads to queue buildup and increased latency during traffic bursts, whereas provisioning for peak demand wastes computational resources during low-load periods~\cite{yu2024moesys}. Autoscaling addresses this trade-off by dynamically adjusting the worker pool orchestration based on application-level signals, such as queue depth and worker utilization, thereby matching resource allocation to workload demand without prior knowledge of future arrivals~\cite{7937885}. A second challenge is model selection. Modern NLP systems employ diverse language models, ranging from zero-shot models such as BART-large-MNLI and DeBERTa-v3-large to fine-tuned models such as DistilBERT~\cite{yang2024give}. These models differ in accuracy, latency, and computational cost, yet model selection is often coupled with infrastructure provisioning. In practice, however, these are independent design dimensions: the autoscaling layer determines the number of active workers, whereas the model determines task-level prediction performance~\cite{cai2024llmaas,jeong2024arascaler}. Therefore, an effective agentic framework should support interchangeable language models without modifying the autoscaling strategy.

To address these challenges, this paper proposes an adaptive agentic autoscaling framework for LLM-driven text classification based on the cloud environment. The framework introduces a classifier-agnostic application-layer autoscaling mechanism that dynamically adjusts computational resources based on workload conditions while allowing different LLM classifiers to be integrated without modifying the scaling logic. The proposed framework employs asynchronous worker orchestration, queue-aware scheduling, and lightweight threshold-based scaling to improve resource utilization and execution efficiency without requiring external orchestration platforms. By separating infrastructure management from classification logic, the framework enables independent optimization of scalability and model quality across different LLM-driven classification workloads. The main contributions of this work are summarized as follows:

\begin{itemize}

\item We propose an adaptive agentic autoscaling framework for cloud-based LLM-driven text classification that explicitly worker-pool orchestration elasticity from language model selection, allowing different classifiers to be integrated without modifying the autoscaling mechanism.

\item We design and implement a lightweight five-component architecture, comprising a priority task queue, dynamically scalable asynchronous worker pool orchestration, centralized metrics collector, threshold-based dynamic autoscaler, and asynchronous orchestrator. 

\item We develop two complementary algorithms that govern the proposed framework: (i) an asynchronous agent worker loop orchestration for concurrent task execution, metric collection, and graceful worker termination, and (ii) a threshold-based autoscaling control loop that dynamically provisions and deprovisions worker instances according to queue depth and worker utilization while respecting configurable scaling bounds and cooldown intervals.

\item We perform a comprehensive experimental evaluation on two benchmark text-classification datasets, comparing the proposed framework with static allocation and standalone classification baselines. The results demonstrate improved resource efficiency, latency, throughput, and scalability while confirming that infrastructure elasticity and model selection are independent design dimensions.

\end{itemize} 

The rest of the paper is organized as follows. Section~\ref{sec:relatedwork} reviews the related work. Section~\ref{systemarchitecture} presents the system architecture of the proposed framework. Section~\ref{sec:systemmodel} introduces the system model and mathematical formulation. Section~\ref{methodology} presents the implementation methodology, including the system workflow and proposed algorithms. Section~\ref{datasetandexperimental} presents the experimental setup and benchmark datasets. Section~\ref{resultsec} discusses the experimental results. Finally, section~\ref{sec:conclusion} concludes the paper and outlines future research directions.

\section{Related Work}
\label{sec:relatedwork}
Recent advances in cloud computing and LLMs have accelerated research in cloud autoscaling, transformer-based text classification, and agentic AI systems. These complementary research directions provide the foundation for scalable intelligent cloud applications. 
\begin{itemize}
    \item Cloud and Container Autoscaling: Autoscaling is a key mechanism for maintaining application performance while efficiently utilizing cloud resources. Reactive approaches, such as Horizontal Pod Autoscaler (HPA)~\cite{10829852,kubernetes}, adjust service replicas using infrastructure-level metrics including CPU and memory utilization. However, HPA often reacts slowly to latency-sensitive AI workloads~\cite{wang2024resource}. To address this limitation, predictive autoscaling techniques based on time-series forecasting and machine learning have been proposed~\cite{pozdniakova2024hpa,abdullah2020burst}. Recent studies further exploit Transformer models for workload prediction and hierarchical decentralized autoscaling~\cite{11190058}, including Kubernetes custom-resource-based autoscaling~\cite{kumar2024optimizing}, multivariate MAPE frameworks~\cite{kumar2025multivariate}, and InformerAutoScale for long-term workload forecasting~\cite{kumar2025optimizing}. For LLM serving such as ARAScaler~\cite{jeong2024arascaler}, LLMaaS~\cite{cai2024llmaas} and recent distributed GPU-cluster serving frameworks~\cite{11143904} optimize request batching, scheduling, pipeline execution, and resource allocation to improve inference efficiency. Unlike these approaches, our framework performs application-layer autoscaling using queue depth and worker utilization while remaining independent of external orchestration platforms.
    \item Transformer-Based Text Classification: Transformer architectures have become the dominant paradigm for text classification. Zero-shot models, such as BART-large-MNLI and DeBERTa-v3, achieve competitive performance without task-specific training~\cite{11079753}, whereas fine-tuned models such as DistilBERT provide an effective balance between inference speed and predictive accuracy~\cite{shah2024fine}. Existing studies primarily evaluate these models in standalone inference settings, emphasizing prediction accuracy rather than execution efficiency. In contrast, our work evaluates multiple classifiers within a unified autoscaling framework, separating classifier performance from infrastructure efficiency.
\item Agentic LLM Systems and Multi-Agent Orchestration:
Recent LLM advances have enabled agentic AI systems in which autonomous agents perform reasoning, planning, memory management, and tool utilization with minimal human intervention~\cite{chowa2026language,wang2024survey,chang2024survey}. Existing research has largely focused on multi-agent collaboration and reasoning, while platforms such as LLMaaS~\cite{cai2024llmaas} and MoESys~\cite{yu2024moesys} improve LLM-serving efficiency through scheduling~\cite{11169423}, batching, and mixture-of-experts routing. However, these approaches generally assume fixed computational resources and do not consider dynamic autoscaling of agent workers. Our framework instead focuses on application-layer worker-pool autoscaling while supporting interchangeable language models through a classifier-agnostic execution layer.
\end{itemize}

\begin{table*}[!ht]
\centering
\caption{Comparison of representative cloud autoscaling and LLM-serving approaches with the proposed framework.}
\label{tab:comparison}
\renewcommand{\arraystretch}{1.25}
\resizebox{\textwidth}{!}{
\begin{tabular}{p{4.6cm}c c c c c c c c}
\hline

\textbf{Approach} &
\textbf{Year} &
\textbf{Predictive Scaling} &
\textbf{Application Layer} &
\textbf{LLM Support} &
\textbf{Classifier Agnostic} &
\textbf{Dynamic Worker Pool} &
\textbf{Cloud Deployment} &
\textbf{NLP Evaluation} \\

\hline

Kubernetes HPA~\cite{10829852,kubernetes}
& 2025
& \xmark
& \xmark
& \xmark
& \checkmark
& \xmark
& \checkmark
& \xmark \\

Predictive Autoscaling~\cite{pozdniakova2024hpa,abdullah2020burst}
& 2024
& \checkmark
& \xmark
& \xmark
& \checkmark
& \xmark
& \checkmark
& \xmark \\

Proactive Scaling with Predictive Models and Custom Resources~\cite{kumar2024optimizing}
& 2024
& \checkmark
& \checkmark
& \xmark
& \checkmark
& \checkmark
& \checkmark
& \xmark \\

Transformer-based MAPE Autoscaling Framework~\cite{kumar2025multivariate}
& 2025
& \checkmark
& \checkmark
& \xmark
& \checkmark
& \checkmark
& \checkmark
& \xmark \\

InformerAutoScale~\cite{kumar2025optimizing}
& 2025
& \checkmark
& \checkmark
& \xmark
& \checkmark
& \checkmark
& \checkmark
& \xmark \\

ARAScaler~\cite{jeong2024arascaler}
& 2024
& \checkmark
& \checkmark
& \checkmark
& \xmark
& \checkmark
& \checkmark
& \xmark \\

LLMaaS~\cite{cai2024llmaas}
& 2024
& \xmark
& \checkmark
& \checkmark
& \xmark
& \checkmark
& \checkmark
& \LEFTcircle \\

MoESys~\cite{yu2024moesys}
& 2024
& \xmark
& \checkmark
& \checkmark
& \xmark
& \xmark
& \checkmark
& \checkmark \\

Standalone Transformer Classifiers~\cite{11079753,shah2024fine}
& 2024-2025
& \xmark
& \xmark
& \checkmark
& \checkmark
& \xmark
& \xmark
& \checkmark \\

Agentic LLM Systems~\cite{chowa2026language,11143904,11169423}
& 2024-2026
& \xmark
& \LEFTcircle
& \checkmark
& \checkmark
& \LEFTcircle
& \LEFTcircle
& \xmark \\

\hline

\textbf{Proposed Framework}
& \textbf{2026}
& \checkmark
& \checkmark
& \checkmark
& \checkmark
& \checkmark
& \checkmark
& \checkmark \\

\hline

\end{tabular}}
\begin{flushleft}
\footnotesize
\textbf{Notation:} $\checkmark$ = Fully addressed; $\LEFTcircle$ = Partially addressed; $\times$ = Not addressed;
\end{flushleft}
\end{table*}

Despite significant progress in these research areas, they remain largely independent. Existing autoscaling approaches optimize infrastructure resources, LLM-serving frameworks focus on inference efficiency, and transformer-based text classification emphasizes predictive performance. Table~\ref{tab:comparison} summarizes representative approaches and highlights that limited attention has been given to unified frameworks combining application-layer autoscaling, classifier-agnostic execution, and dynamic worker-pool management for LLM-driven text classification.

\section{System Architecture} \label{systemarchitecture}

\begin{figure*}[!ht]
    \centering
\includegraphics[width=0.99\linewidth]{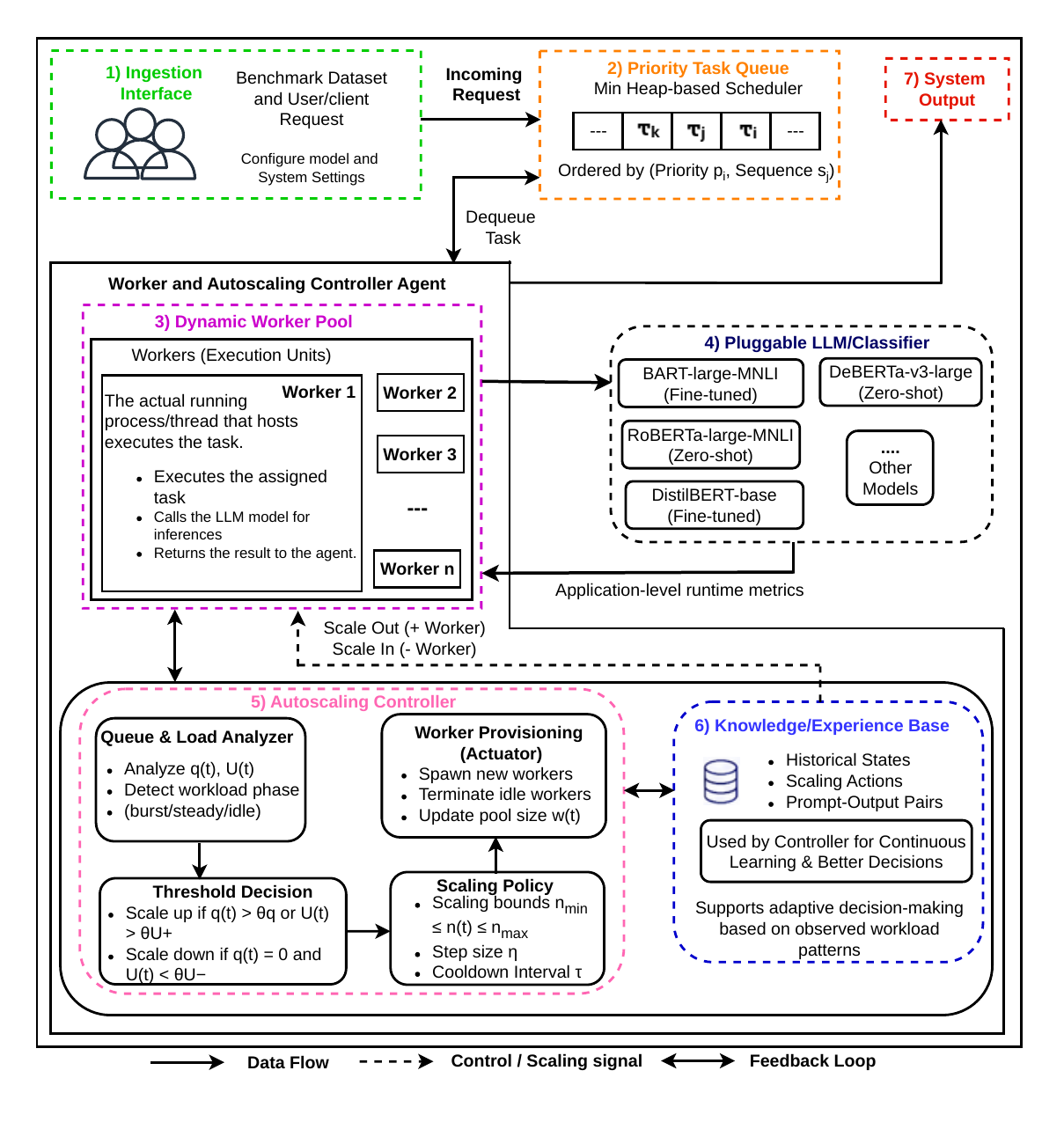}
    \caption{Architecture of the proposed agentic autoscaling framework}
    \label{fig:placeholder}
\end{figure*}

Figure~\ref{fig:placeholder} presents the internal architecture of the proposed agentic autoscaling framework. Incoming benchmark data or user requests are first received through the ingestion interface and converted into tasks submitted to the priority task queue, where tasks are ordered according to priority and sequence number. The queued tasks are then dispatched to the dynamic worker pool, which combines task execution agents with multiple asynchronous workers. The asynchronous workers, worker pool, and autoscaling controller agent collectively constitute the Worker and Autoscaling Controller component. After a task is received from the dynamic worker pool, each task execution agent coordinates its execution with an available worker, invokes the selected LLM/classifier, and tracks the task status and result. The worker acts as the actual execution unit, processing the assigned task asynchronously, invoking the selected model for inference, and returning the result to the agent. The worker pool maintains multiple asynchronous workers and dynamically expands or contracts according to workload demand. Each worker invokes the pluggable LLM/classifier interface, which supports BART-large-MNLI, RoBERTa-large-MNLI, DeBERTa-v3-large, and DistilBERT-base without modifying the worker or autoscaling logic. During execution, the framework continuously monitors application-level runtime conditions, including queue depth, worker utilization, latency, and throughput.
These runtime states are collected by the dynamic worker pool and passed to the autoscaling controller, which contains an autoscaling agent responsible for observing and analyzing the current workload and worker state. The autoscaling agent evaluates queue depth $q(t)$ and worker utilization $U(t)$ together with information maintained in the Knowledge/Experience Base, which stores historical workload states, scaling actions, and observed outcomes. This historical information provides contextual knowledge for assessing previous scaling responses and supporting the autoscaling decision process. Based on the observed state and defined scaling policy, the autoscaling agent determines whether to scale up or scale down. 

It enforces the minimum and maximum worker bounds, scaling step size, and cooldown interval before issuing a scaling action. The resulting decision is communicated back to the dynamic worker pool through the autoscaling control loop. A scale-out action creates additional asynchronous workers when workload demand increases, whereas a scale-in action terminates eligible idle workers when demand decreases, without interrupting active tasks. The resulting worker state and execution outcomes are fed back to the autoscaling agent and the Knowledge/Experience Base, enabling continuous interaction between workload observation, scaling decisions, and system outcomes. Thus, the task execution agents coordinate individual classification tasks, while the autoscaling agent observes workload conditions and manages worker-pool capacity, with the Knowledge/Experience Base providing historical context for the control process. Finally, the framework produces classification results and system-level performance measures, including accuracy, latency, throughput, active workers, resource usage, and resource efficiency.

\section{System Model} \label{sec:systemmodel}
\begin{figure*}
\centering
\includegraphics[width=0.90\linewidth]{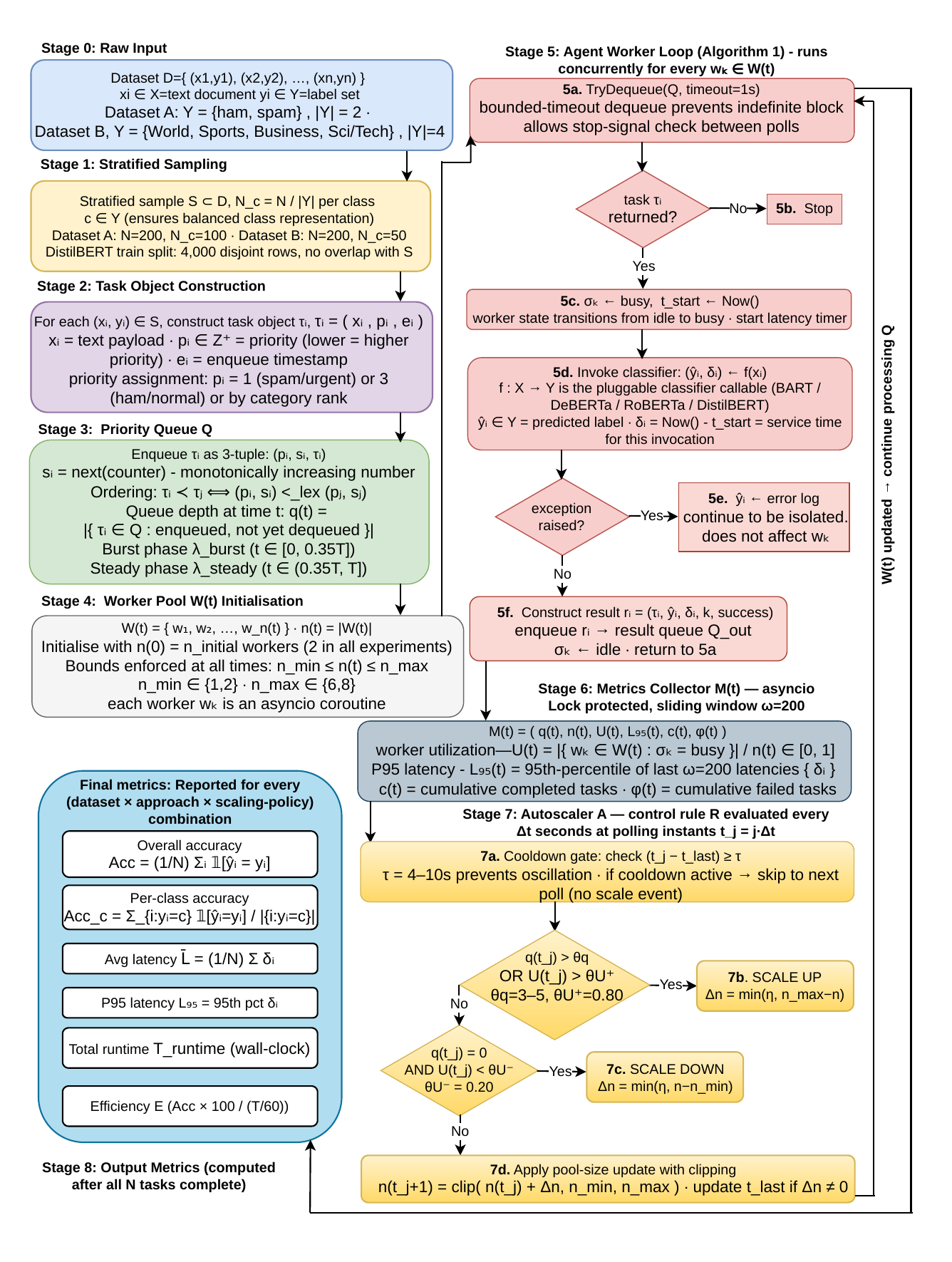}
\caption{End-to-end workflow of the proposed agentic autoscaling pipeline with dynamic worker scaling and performance evaluation}
\label{fig:pipeline-flowchart}
\end{figure*}

Figure~\ref{fig:pipeline-flowchart} illustrates the complete execution flow of the proposed agentic autoscaling framework. Starting from dataset ingestion, the pipeline performs task construction, priority scheduling, asynchronous worker execution, dynamic autoscaling, and performance evaluation. The mathematical formulations introduced in section~\ref{methodology} are implemented through the following eight processing stages.

\begin{itemize}
\item Stage~0 \textendash\ Dataset Input: The pipeline accepts a labeled dataset
$\mathcal{D}=\{(x_i,y_i)\}_{i=1}^{N}$.
Experiments are conducted on two benchmark datasets: AG News for multi-class news classification and SMS Spam Collection for binary spam detection.
\item Stage~1 \textendash\ Stratified Sampling: A balanced evaluation subset is generated using stratified sampling to preserve the original class distribution. The sampled instances are subsequently transformed into executable task objects while maintaining strict separation between training and testing data.
\item Stage~2 \textendash\ Task Construction: Each sampled document is converted into a task object
$\tau_i=(x_i,p_i,e_i)$, where the priority value and enqueue timestamp are assigned before insertion into the scheduling queue.
\item Stage~3 \textendash\ Priority Queue Management: Task objects are inserted into the priority queue according to the ordering defined in section~\ref{methodology}. Incoming requests follow a burst-then-steady workload pattern, allowing the framework to evaluate autoscaling performance under realistic workload fluctuations.

\item Stage~4 \textendash\ Worker Pool Initialization: The worker pool orchestration is initialized with a minimum number of asynchronous workers. Each worker operates independently and continuously retrieves tasks from the shared priority queue while remaining bounded by the minimum and maximum pool-size constraints.
\item Stage~5 \textendash\ Agent Worker Execution: Each worker repeatedly dequeues a task, invokes the selected language model, records the prediction latency, and returns the classification result. Since the worker implementation is independent of the underlying classifier, BART-large-MNLI, DeBERTa-v3-large, RoBERTa-large-MNLI, DistilBERT, or any compatible model can be substituted without modifying the execution logic. Runtime exceptions are isolated at the worker level to prevent failures from propagating to other concurrent workers.
\item Stage~6 \textendash\ Runtime Monitoring: The metrics collector continuously updates queue depth, worker utilization, latency, throughput, and task completion statistics, providing the runtime state required for autoscaling decisions.

\item Stage~7 \textendash\ Dynamic Autoscaling: The autoscaler periodically monitors queue depth and worker utilization. Based on predefined scale-out, scale-in, and cooldown thresholds, it dynamically adjusts the worker pool while ensuring that active tasks are not interrupted. This feedback loop enables the framework to respond efficiently to workload orchestration fluctuations.
\item Stage~8 \textendash\ Performance Evaluation: After all tasks have been processed, the framework computes classification accuracy, per-category accuracy, latency, throughput, and resource-efficiency metrics. 
\end{itemize}
These processing stages collectively implement the proposed agentic autoscaling framework and provide the basis for the experimental evaluation presented in section~\ref{resultsec}.

\section{METHODOLOGY} \label{methodology} 
This section presents the implementation methodology of the proposed agentic autoscaling framework, including the system workflow, asynchronous agent worker algorithm, and threshold-based autoscaling algorithm.

\subsubsection{Task Stream and Arrival Process:}
Let $\mathcal{D} = \{x_1, x_2, \ldots, x_N\}$ denote a dataset of $N$ input documents to be classified, where each $x_i$ is paired with a ground-truth label $y_i \in \mathcal{Y}$ drawn from a label set $\mathcal{Y}$ (for AG News, $\mathcal{Y} = \{\text{World}, \text{Sports}, \text{Business}, \text{Sci/Tech}\}$, $|\mathcal{Y}| = 4$; for the SMS Spam Collection, $\mathcal{Y} = \{\text{ham}, \text{spam}\}$, $|\mathcal{Y}| = 2$). Each document is wrapped in a task object
\[
  \tau_i = (x_i, p_i, e_i)
\]
where $p_i \in \mathbb{Z}^+$ is a priority value (lower values indicate higher priority; in our implementation, priority is assigned by category to allow time-sensitive classes to be processed first) and $e_i$ is the enqueue timestamp. Tasks arrive into the system according to a time-varying arrival process $\lambda(t)$, the instantaneous arrival rate at time $t$. We model arrivals as occurring in two phases to reflect realistic bursty traffic:
\[
  \lambda(t) =
  \begin{cases}
    \lambda_{\text{burst}}, & t \in [0, \beta T] \\
    \lambda_{\text{steady}}, & t \in (\beta T, T]
  \end{cases}
  \qquad \lambda_{\text{burst}} \gg \lambda_{\text{steady}}
\]
where $T$ is the total observation window and $\beta \in (0, 1)$ is the fraction of the window during which burst-phase arrivals occur (in our experiments, $\beta \approx 0.35$--$0.40$, with $\lambda_{\text{burst}}$ approximately 4--7$\times$ larger than $\lambda_{\text{steady}}$,
constructed to stress-test the autoscaler's response to a sudden load spike followed by a return to baseline demand).

\subsubsection{Priority Task Queue:} Tasks are buffered in a priority queue $Q$, implemented as a binary min-heap
ordered first by priority $p_i$ and, to break ties deterministically and avoid undefined comparison behaviour between non-orderable payloads, by a monotonically increasing sequence number $s_i$ assigned at enqueue time:
\[
  \tau_i \prec \tau_j \iff (p_i, s_i) <_{\text{lex}} (p_j, s_j)
\]
We define the instantaneous queue depth as
\[
  q(t) = |\{\tau_i \in Q : \tau_i \text{ enqueued, not yet dequeued at time } t\}|
\]
The queue depth $q(t)$ reflects workload intensity and is used by the autoscaler together with worker utilization.

\subsubsection{Agent worker:} An agent worker $w_k$ is a long-lived asynchronous process that repeatedly executes the loop: dequeue a task $\tau_i$ from $Q$; invoke a classification function $f: \mathcal{X} \to \mathcal{Y}$ on the task payload $x_i$; record the result and elapsed time; enqueue the result. We treat $f$ as an opaque, pluggable callable - the agentic abstraction
at the center of this paper's contribution -- satisfying only the type signature
\[
  f(x_i) \mapsto (\hat{y}_i, \delta_i)
\]
where $\hat{y}_i \in \mathcal{Y}$ is the predicted label and $\delta_i \in \mathbb{R}^+$ is the wall-clock service time for that invocation. In our experiments, $f$ is instantiated as one of: a zero-shot natural-language-inference classifier (BART-large-MNLI, DeBERTa-v3-large,
RoBERTa-large-MNLI), or a supervised fine-tuned classifier (DistilBERT). Crucially, no component of the queue, worker pool, metrics collector, or autoscaler inspects or depends on which $f$ is in use; this is the formal statement of the classifier-agnostic property claimed in Contribution. At any time $t$, each worker is in one of two states, busy or idle:
\[
  \sigma_k(t) \in \{\texttt{busy}, \texttt{idle}\}
\]
A worker is busy for the duration of one $f$ invocation, $\delta_i$ seconds, and idle while waiting on the (possibly empty) queue. 

\subsubsection{Worker pool:} The worker pool at time $t$ is the set of currently active workers,
\[
  W(t) = \{w_1, w_2, \ldots, w_{n(t)}\}, \qquad n(t) = |W(t)|
\]
We refer to $n(t)$ as the pool size. The pool is bounded:
\[
  n_{\min} \le n(t) \le n_{\max} \qquad \forall t
\]
where $n_{\min}, n_{\max} \in \mathbb{Z}^+$ are operator-configured lower and upper bounds (in our experiments, $n_{\min} \in \{1, 2\}$ and $n_{\max} \in \{4, 6, 8\}$ depending on the run). Pool size changes only at discrete control instants determined by the autoscaler (section \ref{methodology}); between control instants, $n(t)$ is constant. We define worker utilization at time $t$ as
\[
  U(t) = \frac{|\{w_k \in W(t) : \sigma_k(t) = \texttt{busy}\}|}{n(t)}
  \in [0, 1]
\]
The worker utilization $U(t)$ reflects processing capacity and, together with queue depth, guides autoscaling decisions. 

\subsubsection{Autoscaling Control Model:} We formalize the autoscaler as a control loop that samples the system state at a fixed polling interval $\Delta t$ and applies a control
function $R$ to determine pool-size adjustments. Let $t_j = j \cdot\Delta t$ for $j = 0, 1, 2, \ldots$ denote the sequence of polling instants. At each $t_j$, the autoscaler observes the state pair
\[
  \mathbf{s}_j = (q(t_j), U(t_j))
\]
and computes a pool-size delta $\Delta n_j \in \mathbb{Z}$ according to the rule

\[
\Delta n_j =
\begin{cases}
+\min(\eta,\, n_{\max}-n(t_j)), & \text{if } C_{\text{up}},\\
-\min(\eta,\, n(t_j)-n_{\min}), & \text{if } C_{\text{down}},\\
0, & \text{otherwise}.
\end{cases}
\]

The auxiliary conditions are defined as

\[
\begin{aligned}
C_{\text{up}} &=
\big(q(t_j)>\theta_q
\lor U(t_j)>\theta_U^{+}\big)
\land (t_j-t_{\text{last}})\ge\tau,\\
C_{\text{down}} &=
\big(q(t_j)=0
\land U(t_j)<\theta_U^{-}\big)
\land (t_j-t_{\text{last}})\ge\tau.
\end{aligned}
\]

Here $\theta_q \in \mathbb{Z}^+$ is the queue-depth scale-up threshold, $\theta_U^{+} \in (0,1)$ and $\theta_U^{-} \in (0,1)$ are the scale-up and scale-down utilization thresholds respectively (with $\theta_U^{-} < \theta_U^{+}$), $\eta \in \mathbb{Z}^+$ is the maximum step size per control instant, $\tau \in \mathbb{R}^+$ is a cooldown period, and $t_{\text{last}}$ is the timestamp of the most recent non-zero $\Delta n$ scaling action. The pool size then evolves as
\[
  n(t_{j+1}) = \mathrm{clip}\big(n(t_j) + \Delta n_j,\ n_{\min},\ n_{\max}\big)
\]
and $t_{\text{last}}$ is updated to $t_j$ whenever $\Delta n_j \neq 0$. The cooldown term $(t_j - t_{\text{last}}) \ge \tau$ is the mechanism that prevents oscillatory scaling -- without it, a pool that scales up in response to a transient queue spike could immediately scale back down at the very next polling instant once that single burst is absorbed, only to
scale up again moments later, a pattern that wastes the overhead of worker spin-up/spin-down without net benefit. In our experiments, $\Delta t \in [2, 5]$ seconds and $\tau \in [4, 10]$ seconds, with $\theta_q \in [3, 5]$, $\theta_U^{+} = 0.80$, $\theta_U^{-} = 0.20$, and
$\eta = 2$ as the default scale-step. Note that $R$ depends only on $(q(t_j), U(t_j))$ -- application-layer signals intrinsic to the task-processing semantics of the pipeline -- and
not on any infrastructure-layer signal such as host CPU or memory utilization. This is the formal distinction from container-level autoscalers (Section \ref{systemarchitecture}): $R$ is well-defined and meaningful even when the underlying workers are bound by GPU computation or network-bound API
calls whose host-level CPU footprint may be misleadingly low. 

\subsubsection{Metrics collector:} The metrics collector $M$ maintains a bounded sliding window of the most recent $\omega$ completed task latencies (we use $\omega = 200$),
\[
  \mathcal{L}(t) = \{\delta_i : \tau_i \text{ completed in } (t-\omega', t]\}
\]
and exposes, at any query time $t$, a snapshot
\[
  M(t) = \big(q(t),\ n(t),\ U(t),\ L_{95}(t),\ c(t),\ \phi(t)\big)
\]
where $L_{95}(t)$ is the 95th-percentile latency over $\mathcal{L}(t)$, $c(t)$ is the cumulative count of successfully completed tasks, and $\phi(t)$ is the cumulative count of failed tasks (task failures are isolated at the worker level: an exception raised during one $f(x_i)$ invocation is caught, logged, and reported as a failed result without terminating the worker or affecting any other in-flight task, which is the basis of the framework's fault-isolation property). $M(t)$ is the single source of truth consumed by the autoscaler at every polling instant and is also the basis for all accuracy, latency, and throughput
statistics reported in section \ref{resultsec}.

\subsubsection{Performance Metrics:}
For a completed run over dataset $\mathcal{D}$, let $\hat{y}_i$ denote the predicted label for input $x_i$ as returned by $f$. We report overall
accuracy
\[
  \mathrm{Acc} = \frac{1}{N} \sum_{i=1}^{N} \mathbb{1}[\hat{y}_i = y_i]
\]
and per-category accuracy for each $c \in \mathcal{Y}$,
\[
  \mathrm{Acc}_c = \frac{\sum_{i : y_i = c} \mathbb{1}[\hat{y}_i = y_i]}
                          {|\{i : y_i = c\}|}
\]
For the binary spam-detection task we additionally report the recall on the minority/positive class explicitly, since, as shown in section \ref{resultsec}, overall accuracy alone can mask a near-complete failure on one class when the dataset is class-balanced but the model's predictions are not: recall on class $c$ is $\mathrm{Acc}_c$ restricted to $c =
\text{spam}$. To compare approaches that trade off accuracy against runtime, we define an accuracy-per-time efficiency score
\[
  E = \frac{\mathrm{Acc} \times 100}{T_{\text{runtime}} / 60}
\]
measured in accuracy-percentage-points per minute of total runtime. $E$ penalizes both low accuracy and long runtime in the same scalar quantity, allowing a fair comparison between, for example, a fast-but-less-accurate zero-shot pipeline and a slow-but-more-accurate fine-tuned standalone model.

\subsection{Agent Worker Algorithm} \label{sec:agentworker}
Algorithm~\ref{alg:worker} describes the execution logic of an asynchronous agent worker within the proposed framework. Each worker repeatedly retrieves a task from the shared priority queue, performs text classification using the selected language model, records execution metrics, and returns the prediction result. The algorithm incorporates timeout control and exception handling to ensure robust execution and fault isolation, allowing multiple workers to operate concurrently without affecting one another. This asynchronous and classifier-agnostic design enables efficient parallel task processing while supporting dynamic worker-pool management under varying workloads. Algorithm~\ref{alg:worker} executes independently for every active worker in the pool. Initially, Lines~1-2 initialize the worker state and repeatedly execute the worker loop until a stop signal is received. Lines~3-6 retrieve the highest-priority task from the shared queue using a bounded timeout; if no task is available, the worker continues waiting without busy-waiting. Lines~7-14 change the worker status to \textit{busy}, record the start time, invoke the selected language model, and handle any execution exceptions through a \texttt{try-catch} mechanism to ensure fault isolation. Lines~15-18 compute the task execution latency, package the prediction result together with execution metadata, and insert the completed task into the output queue. Finally, Lines~19-20 update the runtime metrics and return the worker to the \textit{idle} state, allowing it to process the next task. Since the worker implementation is classifier-agnostic, BART-large-MNLI, DeBERTa-v3-large, RoBERTa-large-MNLI, DistilBERT, or any compatible language model can be integrated without modifying the execution logic. Consequently, infrastructure elasticity and model selection remain independent, enabling different language models to be evaluated under the same autoscaling framework. 

\begin{algorithm}
\caption{Agent worker loop}
\label{alg:worker}
\footnotesize
\renewcommand{\arraystretch}{1.1}
\begin{algorithmic}[1]
\REQUIRE Task queue $Q$, result queue $Q_{\text{out}}$, metrics collector
         $M$, classification function $f$, stop signal $\textsc{Stop}$
\STATE $\sigma_k \leftarrow \texttt{idle}$
\WHILE{$\neg \textsc{Stop}$}
  \STATE $\tau_i \leftarrow \textsc{TryDequeue}(Q, \text{timeout} = 1\text{s})$
  \IF{$\tau_i = \textsc{None}$}
    \STATE \textbf{continue} \COMMENT{re-check stop signal, avoid busy-wait}
  \ENDIF
  \STATE $\sigma_k \leftarrow \texttt{busy}$
  \STATE $t_{\text{start}} \leftarrow \textsc{Now}()$
  \STATE \textbf{Try}
    \STATE $(\hat{y}_i,\_) \gets f(x_i)$
    \STATE $\text{success} \gets \textsc{True}$    
    \STATE \textbf{catch} (exception $\epsilon$)  
    \STATE $\hat{y}_i \gets \textsc{Error}$
    \STATE $\text{success} \gets \textsc{False}$
    \STATE \textsc{Log}$(k,\tau_i,\epsilon)$    
  \STATE \textbf{end try}
  \STATE $\delta_i \leftarrow \textsc{Now}() - t_{\text{start}}$
  \STATE $r_i \leftarrow (\tau_i, \hat{y}_i, \delta_i, k, \text{success})$
  \STATE $\textsc{Enqueue}(Q_{\text{out}}, r_i)$
  \STATE $M.\textsc{Record}(r_i)$
  \STATE $\sigma_k \leftarrow \texttt{idle}$
\ENDWHILE
\end{algorithmic}
\end{algorithm}

\subsection{Autoscaling Algorithm}
\label{sec:autoscaleralgo} 

Algorithm~\ref{alg:autoscaler} presents the threshold-based control strategy used to dynamically manage the agent worker pool orchestration. The autoscaler periodically monitors queue depth and worker utilization to estimate the current workload and processing capacity. Based on predefined upper and lower thresholds, it performs scale-up or scale-down operations while enforcing cooldown intervals to prevent frequent oscillations. Since scaling decisions rely exclusively on application-level metrics rather than infrastructure utilization, the proposed approach provides responsive and efficient resource management for LLM-driven text classification workloads. Algorithm~\ref{alg:autoscaler} executes continuously throughout system operation. Line~1 initializes the autoscaler parameters, including the monitoring interval, cooldown period, scaling thresholds, scaling step size, and worker-pool bounds. Lines~2-5 periodically collect a snapshot of the current system state and skip the scaling cycle if the cooldown interval has not yet expired. Lines~6-10 evaluate the scale-up condition by checking whether the queue depth or worker utilization exceeds the predefined thresholds; if satisfied, additional workers are provisioned while respecting the maximum pool size. Lines~11-15 evaluate the scale-down condition when the queue becomes empty and worker utilization falls below the lower threshold, releasing idle workers without violating the minimum pool size. The control loop then repeats, continuously adapting the worker pool orchestration to workload variations while maintaining stable system operation.
 
\begin{algorithm}
\caption{Autoscaler control loop}
\label{alg:autoscaler}
\footnotesize
\renewcommand{\arraystretch}{1.1}
\begin{algorithmic}[1]
\REQUIRE Metrics collector $M$, worker pool $W$, poll interval $\Delta t$,
         cooldown $\tau$, thresholds $\theta_q, \theta_U^{+}, \theta_U^{-}$,
         step $\eta$, bounds $n_{\min}, n_{\max}$
\STATE $t_{\text{last}} \leftarrow -\infty$
\WHILE{$\neg \textsc{Stop}$}
  \STATE \textsc{Sleep}$(\Delta t)$
  \STATE $(q, n, U, L_{95}, \ldots) \leftarrow M.\textsc{Snapshot}()$
  \IF{$\textsc{Now}() - t_{\text{last}} < \tau$}
    \STATE \textbf{continue} \COMMENT{cooldown active, skip this cycle}
  \ENDIF
  \IF{$q > \theta_q \lor U > \theta_U^{+}$}
    \STATE $\Delta n \leftarrow \min(\eta,\ n_{\max} - n)$
    \IF{$\Delta n > 0$}
      \STATE $W.\textsc{ScaleUp}(\Delta n)$
      \STATE $t_{\text{last}} \leftarrow \textsc{Now}()$
    \ENDIF
  \ELSIF{$q = 0 \land U < \theta_U^{-}$}
    \STATE $\Delta n \leftarrow \min(\eta,\ n - n_{\min})$
    \IF{$\Delta n > 0$}
      \STATE $W.\textsc{ScaleDown}(\Delta n)$
      \STATE $t_{\text{last}} \leftarrow \textsc{Now}()$
    \ENDIF
  \ENDIF
\ENDWHILE
\end{algorithmic}
\end{algorithm}

The proposed methodology establishes a complete framework for LLM-driven agentic text classification by integrating queue-aware task scheduling, asynchronous worker execution, and application-layer autoscaling within a unified cloud architecture. The following section evaluates the proposed framework through extensive experiments on benchmark datasets, comparing classification performance, execution efficiency, scalability, and resource utilization against representative baseline approaches.

\section{Experimental Setup}
\label{datasetandexperimental}
\begin{figure}[]
\centering
\includegraphics[width=0.70\linewidth]{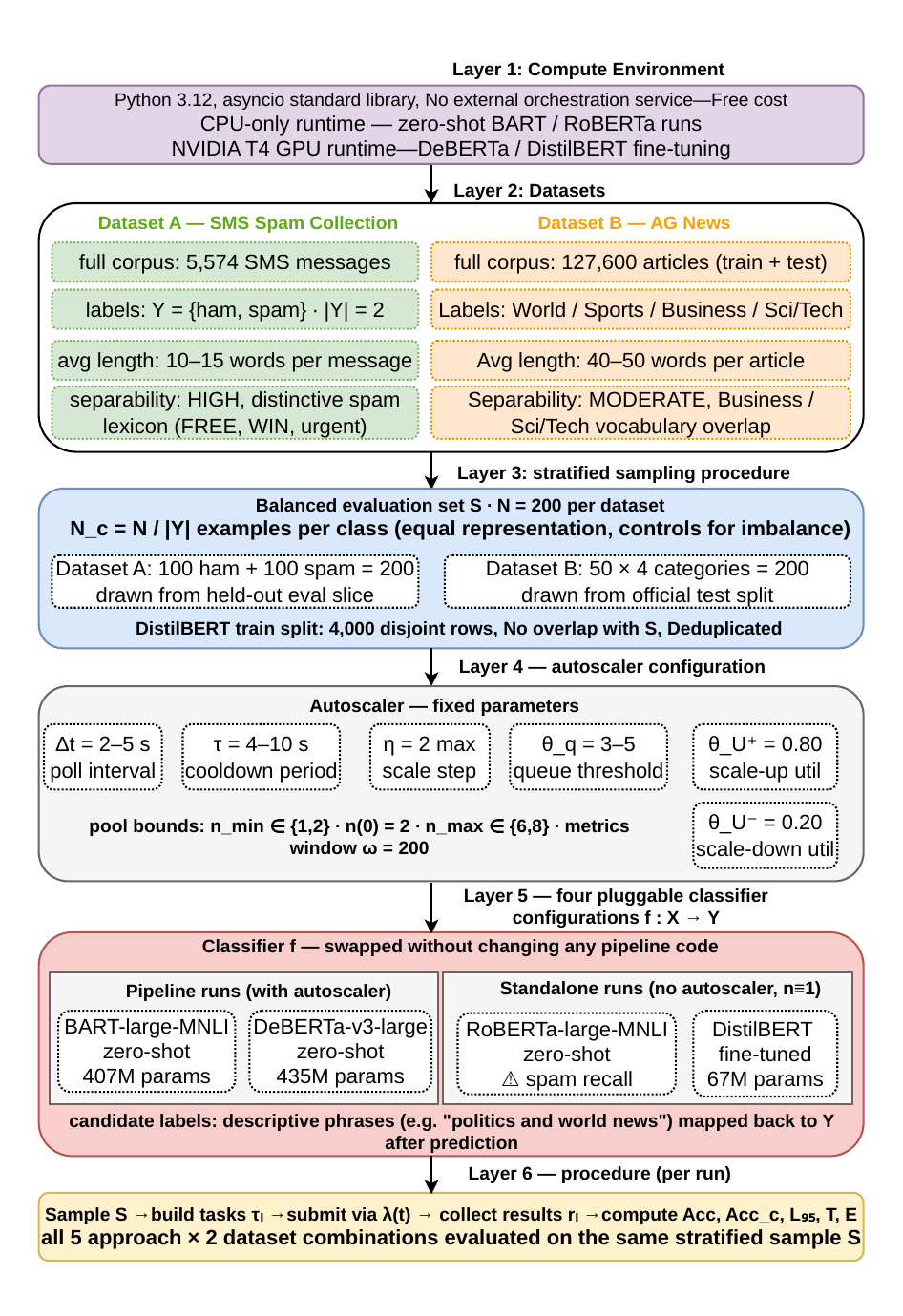}
\caption{Experimental setup of the proposed agentic autoscaling framework}
\label{fig:experiment-setup}
\end{figure}

This section describes the computational environment, software configuration, model integration, dataset preparation, framework initialization, autoscaling configuration, execution procedure, and output generation used to reproduce the experiments. The complete experimental configuration is illustrated in Figure~\ref{fig:experiment-setup}. The procedure is organized as a sequence of practical steps so that the proposed agentic autoscaling framework can be reproduced under the same experimental conditions. 

\begin{enumerate}
    \item \textbf{Compute Environment and Software Setup:} All experiments were conducted using Google Colaboratory as the execution environment. A new Colab notebook was initialized using Python~3.12. The framework uses Python's standard-library \texttt{asyncio} module for asynchronous task execution and does not require Kubernetes, an external task-queue service, or a dedicated cloud orchestration platform. Two runtime configurations were used according to the computational requirements of the evaluated classifiers. BART-large-MNLI and RoBERTa-large-MNLI were executed using the CPU runtime, whereas DeBERTa-v3-large inference and DistilBERT fine-tuning were performed using an NVIDIA T4 GPU runtime. The Colab environment provides approximately 12.7~GB of system memory and session-local temporary storage, while the T4 configuration provides approximately 15~GB of GPU memory. Since Colab resources are session-based and hardware availability may vary, the experiments were performed within a single active runtime for each execution. Reported wall-clock latency therefore corresponds to the native execution environment and is not normalized across CPU and GPU configurations. The required Python libraries were installed before execution using the standard package manager. The main dependencies were \texttt{transformers}, \texttt{torch}, \texttt{datasets}, \texttt{pandas}, and \texttt{accelerate}. The framework itself uses \texttt{asyncio} for worker concurrency and asynchronous orchestration. In Colab, the framework is executed using \texttt{await} because the notebook environment already maintains an active event loop.
    \item \textbf{Language Model Access and Configuration:} The evaluated transformer models were loaded through the Hugging Face \texttt{transformers} library. The models used in the experiments were BART-large-MNLI (\texttt{facebook/ bart-large-mnli}), DeBERTa-v3-large (\texttt{MoritzLaurer/deberta-v3-large-zeroshot-v2.0}), RoBERTa-large-MNLI (\texttt{roberta-large-mnli}), and DistilBERT-base-uncased (\texttt{distilbert-base-uncased}). BART-large-MNLI, DeBERTa-v3-large, and RoBERTa-large-MNLI were used for zero-shot classification, whereas DistilBERT-base-uncased was fine-tuned for supervised classification. The publicly available models can be downloaded directly by the \texttt{transformers} library during the first execution and cached locally within the Colab session. An optional Hugging Face access token can be configured when authenticated access is required; the token is not required for the publicly available models used in this study. For zero-shot classification, the Hugging Face \texttt{pipeline} interface was used with the \texttt{zero-shot-classification} task. CPU execution was specified using \texttt{device=-1}, whereas GPU execution used the corresponding GPU device. Candidate labels were represented using descriptive natural-language phrases rather than only class names. For AG News, the candidate phrases were "politics and world news," "sports and athletics," "business and finance," and "science and technology." For SMS Spam Collection, the candidate phrases were "spam message" and "legitimate message." The candidate phrase receiving the highest NLI entailment score was selected as the prediction and mapped to the corresponding dataset label.

    \item \textbf{Dataset Access and Preparation:} Two benchmark datasets were used. The SMS Spam Collection\footnote{SMS Spam Collection dataset is available at \url{https://github.com/justmarkham/pycon-2016-tutorial/blob/master/data/sms.tsv }}~\cite{Almeida2011Contributions,githubPycon2016tutorialdatasmstsvMaster} dataset was loaded as a tab-separated dataset containing the fields \texttt{label} and \texttt{text}. The dataset contains 5,574 messages, including 4,827 \textit{ham} and 747 \textit{spam} messages. AG News\footnote{AG News dataset is available at \url{https://github.com/mhjabreel/CharCnn_Keras}}~\cite{zhang2015character,githubGitHubMhjabreelCharCnn_Keras} dataset was loaded through the Hugging Face \texttt{datasets} library using its test split. The test split contains 7,600 articles distributed across four categories: World, Sports, Business, and Sci/Tech. To ensure a fair comparison across classifiers and scaling configurations, a balanced stratified evaluation subset of $N=200$ samples was generated independently for each dataset. The number of evaluation samples per class was calculated as
\[
N_c=\frac{N}{|\mathcal{Y}|},
\]
resulting in 100 samples per class for SMS Spam Collection and 50 samples per class for AG News. The same 200 evaluation samples were used for every experimental configuration. For the supervised DistilBERT experiments, an additional disjoint training subset of 4,000 samples was selected from the remaining dataset samples. The training subset was kept separate from the evaluation subset to prevent train--test leakage.

\item \textbf{Framework and Worker Initialization:} After dataset preparation, each evaluation sample was converted into a task object containing the input text, priority information, and enqueue metadata. The proposed framework was initialized through the asynchronous orchestration mechanism with an initial worker count of
\[
n(0)=2.
\]
The worker pool was constrained by predefined minimum and maximum worker limits. Each worker was implemented as an independent asynchronous execution unit and launched through the framework's asynchronous event loop. Workers repeatedly wait for tasks from the shared priority queue, process the assigned task, invoke the selected classification function, and return the prediction and execution statistics. The task execution logic is classifier-agnostic. The classification function is supplied to the worker through a common interface, allowing BART-large-MNLI, DeBERTa-v3-large, RoBERTa-large-MNLI, or DistilBERT to be substituted without modifying the worker implementation or autoscaling policy. Each worker processes tasks independently, and exceptions generated during an individual inference operation are isolated at the worker level so that a failed task does not terminate other concurrent workers.

\item \textbf{Agent and Model Execution:} The task execution agent coordinates the processing of each queued task with an available asynchronous worker. After a task is retrieved from the priority queue, the agent assigns the task for execution and invokes the selected classifier through the common model interface. The worker records the task start time, executes the classifier, captures the prediction and execution status, and returns the result to the agent. For zero-shot models, model inference is executed through the Hugging Face classification pipeline. Because transformer inference can block the asynchronous event loop, the inference operation is dispatched through an executor so that other workers can continue processing tasks concurrently. For DistilBERT, supervised fine-tuning was performed using the Hugging Face \texttt{Trainer} API with mixed-precision training (\texttt{fp16=True}) on the NVIDIA T4 GPU. A batch size of 16 was used, with one training epoch for AG News and two training epochs for SMS Spam Collection. Training and evaluation samples were kept disjoint. After training, the resulting classifier was integrated into the same worker execution interface used by the other models.

\item \textbf{Workload Generation and Task Scheduling:} The 200 evaluation tasks for each dataset were submitted according to the two-phase workload model consisting of a burst-arrival phase followed by a steady-state phase. During the burst phase, tasks were introduced at a rate exceeding the initial processing capacity, producing queue pressure and triggering the autoscaling mechanism. During the subsequent steady-state phase, the arrival rate was reduced, allowing the queue to drain and enabling the autoscaler to release unnecessary workers.

All tasks were inserted into the priority queue and ordered using the priority value and monotonically increasing sequence number. Workers continuously retrieved available tasks from the queue and processed them asynchronously. The same task sequence and evaluation samples were maintained across the compared pipeline configurations to ensure consistent experimental conditions.

\item \textbf{Autoscaling Configuration:} The autoscaler was configured using fixed parameters throughout the pipeline-based experiments. The queue-depth scale-out threshold was set within $\theta_q\in[3,5]$, while the worker-utilization thresholds were set to $\theta_U^{+}=0.80$ for scale-out and $\theta_U^{-}=0.20$ for scale-in. The maximum scaling step was set to $\eta=2$ workers per scaling event. The autoscaler polling interval was configured within $\Delta t\in[2,5]$ seconds, and the cooldown period was set within $\tau\in[4,10]$ seconds. The initial worker pool contained two workers, with $n_{\min}\in[1,2]$ and $n_{\max}\in[6,8]$ depending on the evaluated classifier. At each polling interval, the autoscaling agent obtains the current application state from the runtime metrics and evaluates the queue depth $q(t)$ and worker utilization $U(t)$. A scale-out decision is generated when
\[
q(t)>\theta_q
\quad \lor \quad
U(t)>\theta_U^{+},
\]
whereas scale-in is considered when
\[
q(t)=0
\quad \land \quad
U(t)<\theta_U^{-}.
\]
The scaling step is bounded by $\eta$, and the worker count is maintained within $[n_{\min},n_{\max}]$. A cooldown interval is enforced after each scaling event to prevent repeated or oscillatory scaling actions.

\item \textbf{Runtime Monitoring and Autoscaling Verification:} During execution, the framework continuously records queue depth, active worker count, worker utilization, latency statistics, completed tasks, failed tasks, and throughput. The autoscaling controller periodically obtains a snapshot of these application-level metrics and evaluates the current workload condition. The runtime output includes timestamped status information that allows the scaling behavior to be directly verified. For example, a runtime state may report the current queue depth, number of workers, active workers, utilization, and latency percentile. Scale events are explicitly recorded when workers are added or removed. During the burst-arrival phase, an increase in queue depth or worker utilization should result in scale-out events, whereas after the queue has drained and utilization decreases, eligible idle workers are removed through scale-in events. These runtime logs provide direct evidence that the dynamic worker-pool mechanism is active during the experiment. 

\item \textbf{Knowledge/Experience Interaction:} The Knowledge/Experience Base maintains historical workload states, scaling actions, and observed outcomes associated with previous execution states. The autoscaling agent can access this information as contextual knowledge when evaluating workload behavior and previous scaling responses. The current experimental scaling action remains governed by the predefined threshold-based policy, while the stored experience provides a record of the relationship between observed workload states, scaling actions, and resulting system behavior. The worker state and execution outcomes are subsequently fed back to the Knowledge/Experience Base, maintaining a continuous observation--decision-action cycle. 

\item \textbf{Performance Measurement and Output:} After all $N=200$ evaluation tasks have been completed, the framework aggregates the task-level results and computes the classification and system-performance metrics. Overall accuracy and per-category accuracy are calculated from the predicted and ground-truth labels. Average latency and the 95th-percentile latency are obtained from the recorded task execution durations, while throughput and total execution time quantify processing efficiency. Queue depth and worker utilization are retained to characterize autoscaling behavior, and the resource-efficiency metric is calculated from the achieved classification accuracy and runtime. The framework generates a structured final report containing the classification results and runtime statistics. Task-level results are stored with the task identifier, true label, predicted label, correctness status, execution duration, and worker identifier. The resulting data are subsequently used to generate the performance plots, comparison tables, worker-scaling traces, queue-depth curves, latency and throughput analyses, and resource-efficiency results presented in Section~\ref{resultsec}.

\item \textbf{Baseline and Comparative Evaluation:} The proposed framework was evaluated using Autoscaling+BART and Autoscaling+DeBERTa. These configurations were compared against Static+BART, Standalone RoBERTa, and Standalone DistilBERT. Static+BART uses the same BART-large-MNLI classifier with a fixed two-worker allocation and therefore isolates the effect of dynamic worker management. Standalone RoBERTa performs sequential zero-shot classification without the proposed queue management or autoscaling mechanism, while Standalone DistilBERT uses the fine-tuned model without dynamic worker-pool orchestration. All configurations use the same evaluation samples and consistent performance metrics. This setup enables the effects of classifier selection and worker-pool elasticity to be examined independently.

\end{enumerate}

\section{Results and Discussion}
\label{resultsec}
This section evaluates the proposed agentic autoscaling framework through comprehensive experiments on the AG News and SMS Spam Collection benchmark datasets. The proposed framework is compared with representative static and standalone baselines under identical experimental settings. The discussion is organized into four aspects: (i) dynamic worker scaling and queue evolution, (ii) latency and throughput, (iii) classification performance, and (iv) resource efficiency analysis. Together, these results demonstrate the effectiveness of the proposed application-layer autoscaling framework for scalable LLM-driven text classification.

\begin{figure}[H]
    \centering
    \includegraphics[width=0.90\linewidth]{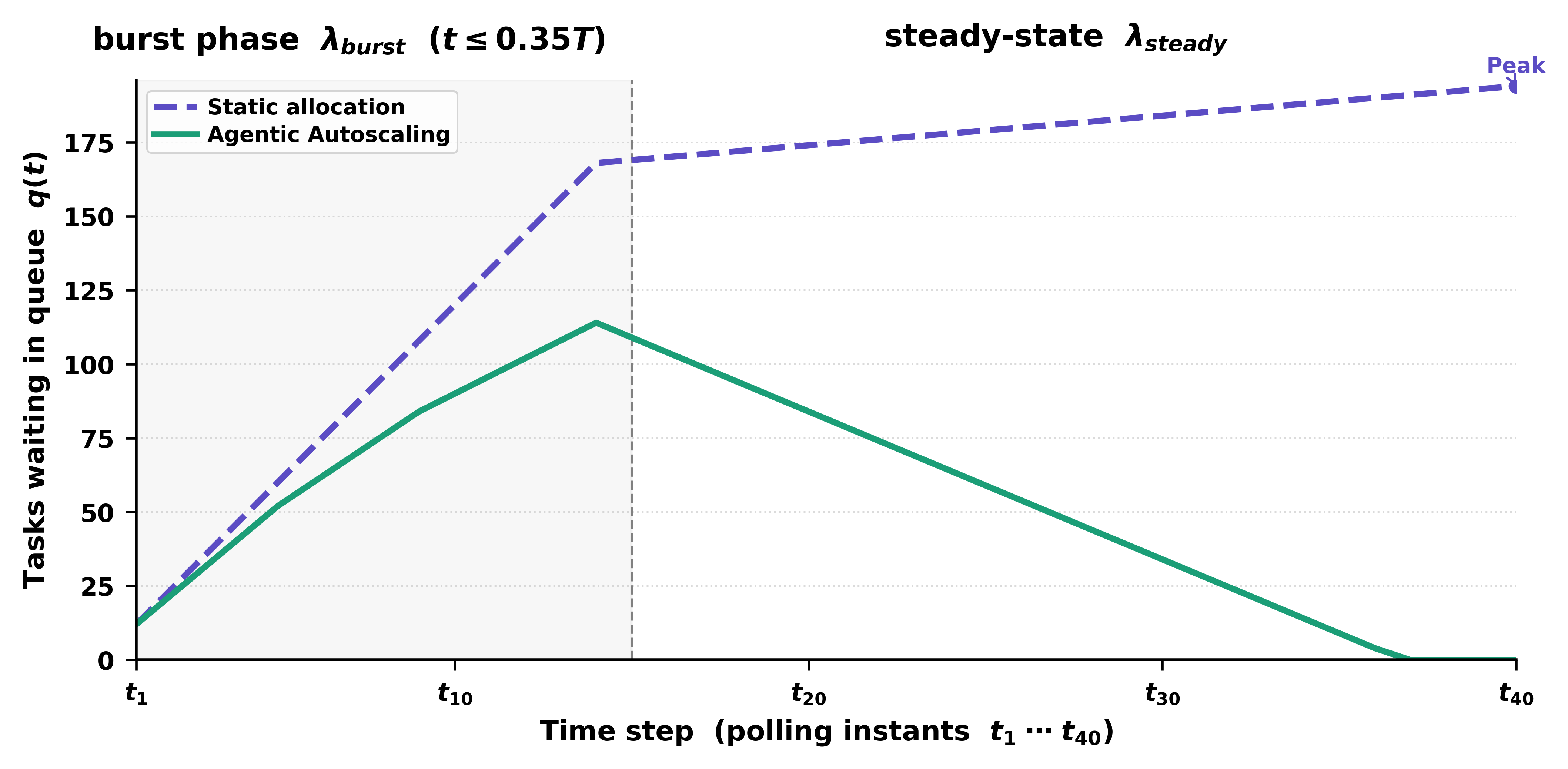}
    \caption{Static worker allocation versus the proposed agentic autoscaling framework under bursty workloads}
    \label{fig:motivation}
\end{figure}

The experimental evaluation is first motivated by the workload behavior illustrated in Figure~\ref{fig:motivation}, which compares static worker allocation with the proposed agentic autoscaling strategy under bursty workloads. With a fixed worker pool, an increase in the task arrival rate beyond the available processing capacity causes queue accumulation, increased waiting time, and higher execution latency, while low-demand periods result in worker underutilization. In contrast, the proposed framework continuously observes application-level conditions, including queue depth and worker utilization, and dynamically adjusts the worker pool according to workload demand. Thus, the motivation for dynamic autoscaling is to increase processing capacity during workload bursts and reduce unnecessary worker allocation during low-demand periods, thereby improving responsiveness and resource efficiency. Let $\lambda(t)$ denote the task arrival rate and let each worker process tasks with an average service rate $\mu$. For a fixed worker pool of size $n$, the processing capacity is $C(t)=n\mu$. Since $n$ remains constant under static allocation, burst periods with $\lambda(t)>C(t)$ cause tasks to arrive faster than they can be processed, resulting in continuous queue growth. The queue dynamics are given by $dq(t)/dt=\lambda(t)-C(t)$, indicating that queue depth increases whenever the arrival rate exceeds the available processing capacity. Consequently, waiting time and end-to-end latency increase, whereas during steady-state operation ($\lambda(t)\ll C(t)$), the fixed worker pool becomes underutilized, wasting computational resources.

To overcome this limitation, the proposed agentic autoscaling framework dynamically adjusts the worker pool according to application-level workload conditions. The processing capacity becomes $C(t)=n(t)\mu$, where $n(t)$ denotes the dynamically allocated worker pool. The autoscaling agent aims to maintain $\lambda(t)\approx C(t)$ by continuously observing queue depth and worker utilization, thereby minimizing queue growth during bursty workloads while avoiding unnecessary resource allocation under light workloads.

\subsection{Dynamic Worker Scaling and Queue Evolution}
\label{sec:dynamic} 
Figures~\ref{fig:agnews-dashboard}(a)--(b) and
\ref{fig:spam-dashboard}(a)-(b) illustrate the runtime behavior of the proposed framework under bursty workloads. In both figures, the x-axis represents the workload observation interval ($t_1$--$t_{40}$), while the y-axis denotes either the number of active workers or the queue depth. Figure~(a) shows that both Autoscaling+BART and Autoscaling+DeBERTa initialize with four workers, expand to six workers at approximately $t_5$, and reach a maximum of eight workers during the burst period ($t_9$--$t_{36}$). As the workload decreases, the worker pool is gradually reduced to seven workers, demonstrating adaptive resource provisioning. In contrast, the Static configuration maintains a constant pool of two workers throughout execution, while the standalone RoBERTa and DistilBERT baselines execute sequentially with a single worker. Figure~(b) presents the corresponding queue-depth evolution. Under the static and standalone configurations, the queue grows continuously throughout the workload, reaching approximately 190-225 pending tasks by the end of execution because the fixed worker pool cannot match the incoming request rate. In contrast, the proposed autoscaling framework limits the queue to approximately 110 tasks during the burst phase and gradually drains it to zero by the end of execution through dynamic worker provisioning. These results demonstrate that application-layer autoscaling effectively matches processing capacity to workload demand, substantially reducing queue buildup and waiting time.
\subsection{Latency and Throughput}
\begin{figure*}
    \centering
    \includegraphics[width=0.99\linewidth]{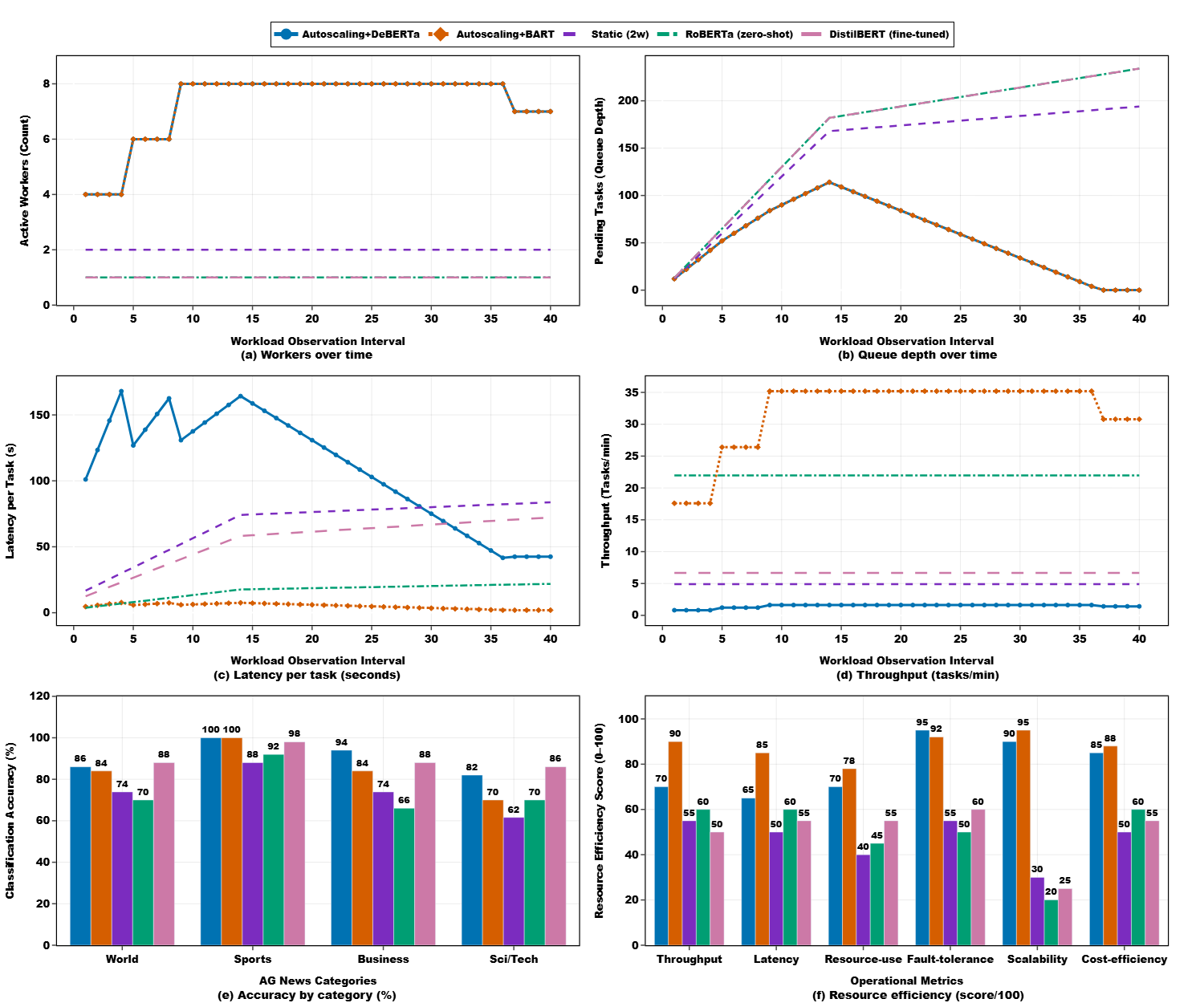}
    \caption{Experimental results on the AG News dataset. The dashboard presents (a) worker scaling, (b) queue depth, (c) latency, (d) throughput, (e) per-category accuracy, and (f) resource efficiency.}
    \label{fig:agnews-dashboard}
\end{figure*}

\begin{figure*}
    \centering
    \includegraphics[width=0.99\linewidth]{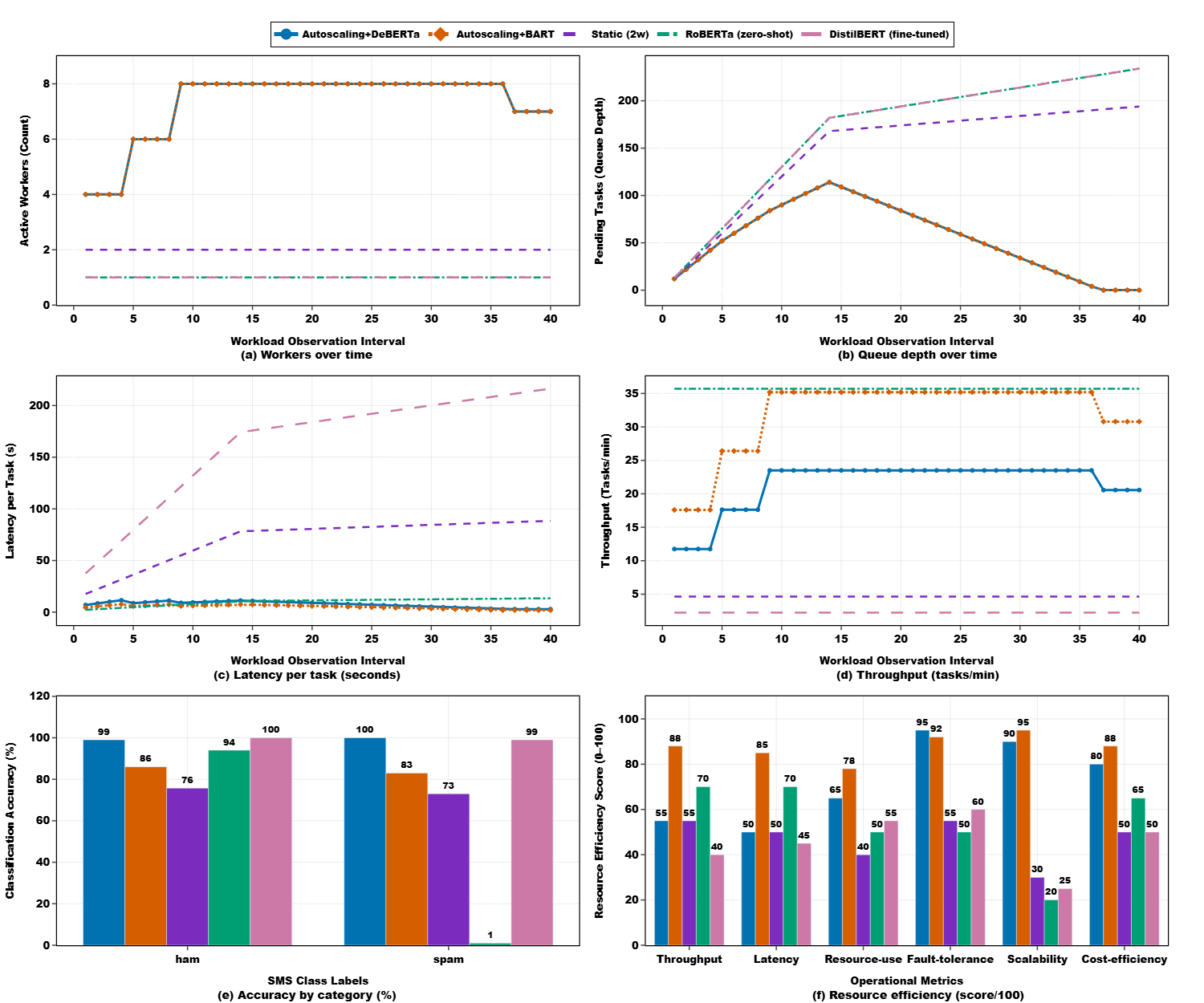}
    \caption{Experimental results on the SMS Spam Collection dataset. The dashboard presents (a) worker scaling, (b) queue depth, (c) latency, (d) throughput, (e) per-label accuracy, and (f) resource efficiency.}
    \label{fig:spam-dashboard}
\end{figure*}

Figures~\ref{fig:agnews-dashboard}(c)--(d) and
\ref{fig:spam-dashboard}(c)-(d) compare the execution latency and throughput of the evaluated approaches. In both subfigures, the x-axis represents the workload observation interval, while the y-axis corresponds to latency per task (seconds) and throughput (tasks/min), respectively. For the AG News dataset, Autoscaling+BART consistently maintains the lowest latency of approximately 5--8\,s throughout execution, whereas Autoscaling+DeBERTa initially exhibits higher latency (approximately 160\,s) owing to the computational complexity of DeBERTa-v3-large before gradually decreasing to about 40\,s as additional workers are provisioned. The Static configuration continues to experience increasing latency, reaching approximately 80\,s by the end of execution because of persistent queue buildup. Similar behavior is observed on the SMS Spam dataset, where Autoscaling+BART maintains latency below 10\,s, while the Static and DistilBERT baselines increase to approximately 90\,s and 220\,s, respectively. Figure~(d) shows that Autoscaling+BART achieves the highest throughput, increasing from approximately 18 tasks/min to 35 tasks/min during the burst period before stabilizing around 31 tasks/min. Autoscaling+DeBERTa achieves approximately 24 tasks/min on AG News and about 1.5 tasks/min on SMS Spam because of the higher inference cost of the larger language model. In contrast, the static and standalone approaches maintain nearly constant throughput throughout execution, indicating their inability to adapt to workload fluctuations. Overall, dynamic worker provisioning significantly improves execution efficiency by increasing parallelism while reducing waiting time. 

\subsection{Classification Performance}
\label{sec:classification} This subsection evaluates the classification effectiveness of the proposed framework on the AG News and SMS Spam Collection datasets. The analysis focuses on overall and per-category classification accuracy while comparing the proposed autoscaling framework with static and standalone baseline configurations.

\subsubsection{AG News Dataset}

Figure.~\ref{fig:agnews-dashboard}(e) summarize the classification results on the AG News dataset. Using the same BART-large-MNLI classifier, the proposed autoscaling framework improves overall accuracy from 75.2\% under static allocation to 84.5\%, while simultaneously reducing execution time through dynamic worker management. Replacing BART with DeBERTa-v3-large further increases overall accuracy to 90.5\% without modifying the queue management, worker execution, or autoscaling strategy, demonstrating that the proposed framework remains independent of the underlying language model. The per-category results show that Autoscaling+DeBERTa consistently achieves the highest accuracy across all four news categories, followed by Autoscaling+BART. In comparison, the static BART configuration exhibits reduced performance, while the standalone DistilBERT and RoBERTa baselines achieve lower overall accuracy. These results indicate that the proposed application-layer autoscaling framework preserves or improves classification performance while enabling interchangeable language models within the same execution environment.

\subsubsection{SMS Spam Collection} 
Figure~\ref{fig:spam-dashboard}(e) present the classification results for the SMS Spam Collection dataset. Owing to the strong lexical distinction between legitimate and spam messages, most evaluated approaches achieve relatively high overall accuracy. Among all methods, Autoscaling+DeBERTa achieves the highest overall accuracy of 99.5\%, whereas Autoscaling+BART attains 84.5\% while providing substantially lower execution time.
A notable observation is the poor performance of the standalone RoBERTa classifier, which achieves only 47.5\% overall accuracy because of extremely low spam-class recall (1\%), despite correctly classifying most ham messages. This result highlights a limitation of NLI-based zero-shot classification for imbalanced binary classification tasks and emphasizes that classifier selection remains critical even when efficient autoscaling is employed. Overall, the proposed framework consistently matches or exceeds the classification performance of the corresponding static configurations while maintaining complete classifier independence. 

\subsection{Resource Efficiency Analysis}
\label{sec:efficiency} 
Figures~\ref{fig:agnews-dashboard}(f) and \ref{fig:spam-dashboard}(f) compare the overall resource efficiency of the evaluated approaches using six complementary dimensions: throughput, latency, resource use (utilization), fault tolerance, scalability, and cost efficiency. On the AG News dataset, Autoscaling+BART achieves the strongest overall efficiency profile, with scores of approximately (90, 85, 78, 92, 95, and 88), while Autoscaling+DeBERTa attains the second-best efficiency profile (70, 65, 70, 95, 90, and 85) together with the highest classification accuracy. In contrast, the static BART configuration exhibits noticeably lower scalability and cost efficiency because its fixed worker pool cannot adapt to workload fluctuations, whereas the standalone RoBERTa and DistilBERT baselines achieve the lowest overall efficiency due to their sequential execution model. A similar trend is observed on the SMS Spam Collection dataset. Autoscaling+BART again provides the best balance across throughput, latency, resource utilization, fault tolerance, scalability, and cost efficiency, with scores of approximately (88, 85, 78, 92, 95, and 88). Autoscaling+DeBERTa achieves the second-best efficiency while delivering the highest classification accuracy, with slightly lower efficiency attributable to its higher computational cost. Both autoscaling configurations consistently outperform the static and standalone baselines across the evaluated resource-efficiency dimensions. 

Overall, the observed changes in worker count and queue depth demonstrate that the proposed agentic worker-pool orchestration can adapt processing capacity to bursty workloads, while the latency and throughput results confirm the effectiveness of asynchronous execution and application-layer autoscaling. The classification results further demonstrate the classifier-agnostic nature of the framework, as different zero-shot and fine-tuned language models can be integrated without modifying the queue management, worker execution, or autoscaling mechanism. At the same time, the comparative results indicate that model selection remains an important factor in determining the accuracy-efficiency trade-off, while the autoscaling mechanism independently manages processing capacity according to workload demand. The resource-efficiency results further show that dynamic worker-pool management provides improved scalability and resource utilization compared with fixed or sequential execution. Collectively, these findings validate the proposed five-component architecture and its two complementary control mechanisms, asynchronous agent-worker execution and threshold-based autoscaling-and demonstrate that worker-pool elasticity and language-model selection can be treated as independent design dimensions for scalable LLM-driven text classification.

\section{Conclusions and Future Work}
\label{sec:conclusion}
This paper proposed an agentic autoscaling framework for LLM-driven text classification that decouples infrastructure elasticity from model selection through a classifier-agnostic worker-pool orchestration architecture implemented entirely using python's \texttt{asyncio} library. Our framework integrates priority-based task scheduling, asynchronous agent execution, runtime monitoring, and application-layer autoscaling without requiring external orchestration platforms. Experimental evaluation on the AG News and SMS Spam Collection benchmark datasets demonstrates that Autoscaling+DeBERTa achieves the highest classification accuracies of 90.5\% and 99.5\%, respectively, while Autoscaling+BART provides a favorable balance between accuracy (84.5\% on both datasets) and execution efficiency. Compared with static resource allocation and standalone baselines, the proposed framework consistently improves latency, throughput, scalability, and overall resource efficiency. The results further demonstrated improvements in queue management, latency, throughput, scalability, and resource efficiency compared with static and standalone baselines. These findings validate the proposed architecture and demonstrate that worker-pool elasticity can be managed independently of language-model selection, enabling different zero-shot and fine-tuned classifiers to operate within a common autoscaling framework.

The current study is limited to two benchmark datasets, manually configured threshold parameters, and evaluation within a single computing environment. Future work will extend the framework to larger and more diverse workloads and additional language models, while investigating automated parameter optimization and predictive autoscaling to anticipate workload changes before queue buildup occurs. Further research will also investigate distributed cloud-edge deployment, intelligent task placement, and heterogeneous resource allocation, enabling lightweight models to operate at the edge while computationally intensive LLMs are dynamically assigned to cloud resources. These extensions can further improve scalability, adaptability, and resource efficiency for large-scale agentic AI applications.

\begin{acks}
We gratefully acknowledge the support of the Institutions of Eminence (IoE) Scheme of Banaras Hindu University, Varanasi, India under Development Scheme No. 6031 and an Australian Research Council (ARC) Discovery Project grant.
\end{acks}

%
\bibliographystyle{ACM-Reference-Format}
\bibliography{sample-base}

\end{document}